\documentclass[]{aastex701}
\usepackage{url}
\usepackage{ulem}
\usepackage[utf8]{inputenc}

\shorttitle{A Radio-Bright Local Little Red Dot Analog}
\begin{document}

\title{A Radio-Bright Local Little Red Dot Analog}

\author[0000-0003-2737-5681]{Luis F. Rodr\'{\i}guez}
\affiliation{Instituto de Radioastronom\'{\i}a y Astrof\'{\i}sica\\
Universidad Nacional Aut\'onoma de M\'exico, Apdo. Postal 3-72, Morelia, Michoac\'an 58089, Mexico}
\email[show]{l.rodriguez@irya.unam.mx}
\author[0000-0002-3210-6307]{I. F\'elix Mirabel}
%\affiliation{D\'epartement d’Astrophysique-IRFU-CEA \\ Universit\'e Paris-Saclay, France}
\affiliation{Instituto de Astronom\'{\i}a y F\'{\i}sica del Espacio (IAFE)\\
CONICET-Universidad de Buenos Aires, C1428 Buenos Aires, Argentina}
\email[show]{mirabelfelix@gmail.com}
\author[0000-0003-1557-4931]{Rosa A. Gonz\'alez-L\'opezlira}
\affiliation{Instituto de Radioastronom\'{\i}a y Astrof\'{\i}sica\\
Universidad Nacional Aut\'onoma de M\'exico, Apdo. Postal 3-72, Morelia, Michoac\'an 58089, Mexico}
\email[show]{r.gonzalez@irya.unam.mx}
\author[0000-0002-5635-3345]{Laurent Loinard}
\affiliation{Instituto de Radioastronom\'{\i}a y Astrof\'{\i}sica\\
Universidad Nacional Aut\'onoma de M\'exico, Apdo. Postal 3-72, Morelia, Michoac\'an 58089, Mexico}
\email[show]{l.loinard@irya.unam.mx}

%\collaboration{all}{The Terra Mater collaboration}

%% Use the \collaboration command to identify collaborations. This command
%% takes an optional argument that is either a number or the word "all"
%% which tells the compiler how many of the authors above the command to
%% show. For example "\collaboration[all]{(DELVE Collaboration)}" wil include
%% all the authors above this command.
%%
%% Mark off the abstract in the ``abstract'' environment. 
\begin{abstract}

The James Webb Space Telescope revealed the existence in the early Universe ($z >$ 4) of large populations of little red dots (LRDs), compact red luminous sources that host rapidly growing supermassive black holes (SMBHs) surrounded by coeval nuclear starbursts. LRDs have defining 
properties, one of which is the absence of radio detections. LRDs could be radio-undetected because of their large distances. A way to investigate the
radio properties of LRDs 
is to search for radio emission in their local analogs,
the Local Little Red Dot (LLRD) galaxies at $z <$ 1 with analogous properties to LRDs. We report the radio continuum detection of the LLRD analog J204837.26${-}$002437.2 ($z$ = 0.4332, hereinafter J2048). This adds to the previous radio detection of two other LLRDs. However, with a flux density of 3.0$\pm$0.2 mJy at 3.0 GHz, J2048 is two orders of magnitude more radio luminous than the previous detections. The spectral index of $\alpha$ = -0.39$\pm$0.04 is consistent with moderately optically-thick synchrotron emission. The radio luminosity of J2048 is $1.2 \times 10^{41}~ \mathrm{erg ~s^{-1}}$, in the lower end of the range defined by radio-loud giant elliptical galaxies and quasars of $L_R$ = $10^{41-46} ~\mathrm{erg ~s^{-1}}$. 
The expected radio luminosity of the SMBH associated with J2048 is estimated to be about an order of magnitude larger,
suggesting that its radio luminosity could potentially be strongly dimmed.
A cosmological ($z >$ 4) LRD with radio luminosity similar to that of J2048 would be detectable using the VLA with moderately long integrations.

\end{abstract}

%% Keywords should appear after the \end{abstract} command. 
%% The AAS Journals now uses Unified Astronomy Thesaurus (UAT) concepts:
%% https://astrothesaurus.org
%% You will be asked to selected these concepts during the submission process
%% but this old "keyword" functionality is maintained in case authors want
%% to include these concepts in their preprints.
%%
%% You can use the \uat command to link your UAT concepts back its source.

%% From the front matter, we move on to the body of the paper.
%% Sections are demarcated by \section and \subsection, respectively.
%% Observe the use of the LaTeX \label
%% command after the \subsection to give a symbolic KEY to the
%% subsection for cross-referencing in a \ref command.
%% You can use LaTeX's \ref and \label commands to keep track of
%% cross-references to sections, equations, tables, and figures.
%% That way, if you change the order of any elements, LaTeX will
%% automatically renumber them.

\section{Introduction} \label{sec:intro}

Several James Webb Space Telescope (JWST)
surveys have discovered unexpected large populations of “Little Red Dots” (LRDs) at redshifts $z>$ 4. These are extremely compact sources
with average effective radius of less than a few hundred pc, widths of broad emission lines of up to 2000 km s$^{-1}$,
and V-shaped spectral energy distributions \citep[SEDs; e.g. ][]{2024ApJ...963..129M, 2025ApJ...986..126K, 2025ApJ...978...92L, 2025ApJ...991...37A}. LRDs reach a maximum density at $z \simeq$ 5 and show an exponential decline in number at redshifts $z<$ 4 \citep{2026A&A...706A..29B}. This implies that local little red dot (LLRD) candidates are scarce.

The nature of the energy source that powers LRDs has been intensively debated since their discovery. As for quasars and ultra-luminous infrared galaxies (ULIRGs) in the past, the possible energy sources \rm have been gravitational energy released by accreting supermassive (SMBHs) or intermediate mass black holes (IMBHs) and/or nuclear energy from
exceptionally intense massive star formation. Several models have addressed this question through the analysis of the blue excess in the rest-frame ultraviolet (UV) SEDs of
LRDs \citep{2025ApJ...988L..22I, 2025arXiv251203130I}.
These authors conclude that LRDs can be interpreted as a transient phase in early SMBH growth and that
they may represent the long-sought missing link
between SMBH seed formation and the emergence of luminous quasars.

%(Inayoshi et al. 2025 https://doi.org/10.48550/arXiv.2509.19422; Inayoshi \& Ho 2025 https://doi.org/10.48550/arXiv.2512.03130).

If LRDs at cosmic dawn are indeed rapidly growing BHs, it is puzzling that they are not detected or are very dim in soft X-rays ($<$ 10 keV) and that
most of them have not been detected in radio wavelengths. The absence of X-rays has been explained by cold gas absorption in
BH atmospheres with densities $n_H \ge 10^9$ cm$^{-3}$, column densities $N_H  \ge 10^{24}$ cm$^{-2}$, and turbulent velocities of
$\sim$500 km s$^{-1}$ \citep{2025MNRAS.538.1921M, 2026ApJ...997..364L}.

However, contrary to X-rays, radio emission from accreting BHs may not be totally or even partially absorbed by such column densities of cold neutral gas. In fact, radio jets from accreting stellar mass BHs are easily observed, even when the soft X-rays are completely 
obscured \citep[][and references therein]{2025ApJ...986..108R}. If radio emission could be detected in LRDs, it may be possible to discriminate from
the morphology, time behavior and spectral index of the emission, 
between radio feedback from BHs and that from massive stars, i.e., between the energy generated by highly accreting IMBH/SMBHs, and that produced by multiple supernovae from rich clusters of massive stars.
Besides, the detailed study of the radio emission
may provide information to determine the specific phase in the evolution of the LRD studied, from
its birth at cosmic dawn, believed to be dominated by feedback from a rapidly growing IMBH, to an intermediate (U)LIRG stage,
to the possibly final post-LRD phase of a galaxy, which may end up with a dormant SMBH surrounded by an extended stellar component \citep{2026A&A...706A..29B}.

LRDs, usually detected in the near-IR in the early universe, so far have been radio-undetected for unknown reasons,
possibly influenced by
their large distances and the limitations in our observational capabilities  
\citep[e.g. ][]{2026A&A...706A.372M, 2025A&A...693L...2P, 2025A&A...694L..14L}.
Several models have attempted to explain the absence of radio emission by invoking free-free absorption, synchrotron self-absorption, or the disruption of magnetic field coronae by super-Eddington accretion  \citep{2026A&A...706A.372M, 2025A&A...701A.168D}.

Besides these models, a tentative cooling mechanism of radio BH-jets that 
to our knowledge has not yet been considered is  Compton scattering, 
by UV photons from massive stars, of the highly relativistic electrons that produce the synchrotron emission of the BH-jets. For instance, in the case of BH-X-ray binaries (XRBs), where the donor star is very luminous, as for example in Cygnus X-1, this dimming mechanism of the radio jets is very important, of several orders of magnitude
\citep{2022NewAR..9401642M}. This inverse Compton (IC) cooling by photons from massive stars affects the microscopic motions of the radiating ultra-relativistic electrons, suppressing the synchrotron radiation. This mechanism, however, does not affect the bulk motion of the ejecta. This is probably why LRDs and other types of active galactic nuclei (AGN) galaxies with nuclear starbursts may be capable of producing energetic mechanical feedbacks with radio “dark” jets. If BH-jets in LRDs and AGN in general co-exist with nuclear clusters of massive stars, the BH-jets may be radio weak but still produce massive outflows. Another radiation field that may
produce inverse Compton weakening in the relativistic electrons
of the jet is the 
cosmic microwave background, since its energy density scales 
as $(1 + z)^4$ \citep{2015MNRAS.452.3457G}.

Despite those dimming mechanisms, BH-jets may not be completely radio “dark”. The black hole activity in the nuclei of galaxies may always be revealed
by the presence of a compact radio-jet or an unresolved radio core \citep{2004A&A...414..895F}.
Therefore, assuming that radio emission from LRDs at high redshifts may not have been detected because they reside at large distances and/or the observational capabilities are limited, we have initiated a search for radio emission from LLRDs with the U.S. National Radio Astronomy Observatory (NRAO) Very Large Array (VLA) and Very Long Baseline Array (VLBA), with the expectation that it might help understand why 
LRDs at high redshifts are radio-undetected, eventually leading to strategies (such as the determination of an optimal frequency),
that could result in the radio detection of selected cosmological
counterparts. An immediate way to test the different LRD models directly is by analyzing archived radio observations of LLRD candidates.

Using data from both the VLA archives and the Very Large Array Sky Survey (VLASS), radio emission was detected associated with two LLDRs: J1025+1402 and J1047+0739, at redshifts $z =$ 0.1$-$0.2 \citep{2026A&A...707L..17R}. 
J1047+0739 is relatively nearby ($z =$ 0.1682), radio faint, but clearly detected (117$\pm$8 
$\mu$Jy at 6.0 GHz).
The radio emission associated with J1025+1402
($z =$ 0.10067) is fainter (45$\pm$10 $\mu$Jy at 6.0 GHz) than that associated with J1047+0739. Furthermore, 
the radio emission is statistically significant but 
displaced by $\sim$2$''$ from the optical position.
Two LRDs apparently in transition to QSOs,  Forge I ($z$ = 2.871) and Forge II ($z$= 2.930), have also been detected in the radio 
continuum \citep{2026NatAs.tmp..118F}. 
Finally, the AGN candidate PRIMER-COS 3866 at $z$ = 4.66 was detected in the radio survey of \citet{2025ApJ...986..130G}. 

Using a search method of optical-radio cross-matching we have seeked for radio emission in 48 LLRDs reported in the papers by \citet{2026ApJ...999...30C}, \citet{2026A&A...709L..11D},
\citet{2025ApJ...980L..34L}, \citet{2026ApJ...997..364L}, \citet{2026arXiv260514233P}, and \citet{2026arXiv260604712S}.
A total of 14 of these sources were detected in the stacked VLASS images, for a success rate of 29\%. 
Here, we report radio observations of the brightest of these sources:
LLRD J204837.26-002437.2 \citep[hereinafter J2048,
as labeled by][]{2026ApJ...999...30C}, at $z =$ 0.4332.
%J2048 hosts an overmassive BH with an extended starburst
%\citep{2026ApJ...999...30C}. 
It is a very interesting source, 
since it is comparatively very radio bright 
(3.1$\pm$0.1 mJy at 3.0 GHz,
as we will see below), and an obvious target for VLBA follow-up observations. In Section 2, 
we present the properties of J2048, and in Section 3 the survey radio observations are
described. We discuss how the radio parameters, in particular 
its radio luminosity, constrain the nature
of J2048 in Section 4. 
The concept of supermassive stars, a possible explanation
for the LRDs, and how they could be weak radio emitters
is summarized in Section 5.
Finally, a discussion of J2048 in the
multiwavelength context is presented in Section 6 and our conclusions are
outlined in Section 7.

\section{The Galaxy J2048}

J2048 is one of the lowest-$z$ analogs of LRDs, at $z =$ 0.4332. It was identified by \citet{2026ApJ...999...30C}, using new Gemini-North Multi-Object Spectrograph (GMOS) Integral Field Unit (IFU) spectroscopic observations, combined with archival multiband photometric SED data. The GMOS data reveal extended blue emission from a starburst with a star formation rate (SFR) of 400 $\mathrm{M_\odot}$ yr$^{-1}$, with a high BH–to–stellar mass ratio of 60\% 
and an extended, highly fast, ionized outflow. J2048 exhibits features similar to those of the high-$z$ LRDs, such as a V-shaped SED in the UV and optical bands, as well as a compact, red continuum component that probably arises from a reddened AGN.
It is located at intermediate redshifts (0.1 $< z <$ 1.0) and was selected by \citet{2020ApJ...900...51C},
from the cross-matched all-sky survey catalogs AKARI (far-IR) and Wide-field Infrared Survey Explorer (WISE; NIR and MIR), and the spectroscopic Sloan Digital Sky Survey (SDSS, optical) catalog. 
%\citet{2020MNRAS.499.2245C} from the Sloan Digital Sky Survey and the Dark Energy Survey Supernova fields}. 
A compact red continuum and a broad H$\alpha$ line from
the AGN broad line region (BLR) were newly detected in the Gemini/GMOS observation of J2048.

The blue-excess is spatially extended and emitted from its intense starburst that is estimated has been taking
place over the recent past ($\sim$20 Myr).
A compact red continuum in the NIR, and the BLR H$\alpha$ line indicate an obscured AGN with $A_V$ = 6.3, which corresponds to a bolometric luminosity of 10$^{13.6}~\mathrm{L_\odot}$. 
This high luminosity is consistent with J2048 being an ULIRG \citep{1996ARA&A..34..749S}.
J2048 also exhibits extended narrow line
emission lines \citep{2026ApJ...999...30C}. It is believed that these extended narrow lines are ionized by stellar light of young stars with 
the BPT diagnostic,
used to identify the dominant source of ionization in emission-line galaxies
\citep{1981PASP...93....5B, 2001ApJ...556..121K}.
The corresponding instantaneous SFR estimated using the extinction-corrected fluxes of the narrow H$\alpha$ or [O II] lines are 140 $\pm$ 10 or 100 $\pm$ 20 $\mathrm{M_\odot~ yr^{-1}}$, respectively, utilizing the empirical functions of \citet{2013seg..book..419C}  and
\citet{2009ApJ...703.1672K}.
The SMBH of J2048 is extremely overmassive. 
\citet{2026ApJ...999...30C} estimate the BH mass from
the AGN optical luminosity and the line width of the 
BLR H$\alpha$. Following the empirical relation of 
\citet{2012ApJ...753..125S} they derive $\mathrm{M_{BH} = 10^{10.2}~ M_\odot}$.
They also obtain a stellar mass of $\mathrm{M_* = 10^{10.4}~ M_\odot}$, by fitting the photometric SED
with a stellar population synthesis model. Then, \citet{2026ApJ...999...30C} derive a BH to stellar mass ratio, $\mathrm{M_{BH}/M_*}$, of $\simeq$0.6, which is approximately two
orders of magnitude larger than the ratio for local E/S0-type galaxies with a similar stellar mass $\mathrm{M_*}$ \citep{2015ApJ...813...82R}.

J2048 possesses a fast ionized outflow, which is revealed
by the blueshifted, broad wings in profiles of all emission lines
in the integrated spectrum \citep{2026ApJ...999...30C}. The outflow wings
dominate the line profiles of the [O III] 4959, 5007 \AA ~doublet with
a flux fraction of 84\%. The [O III] outflow has two
components, a primary wing, with $\mathrm{V_s}$ of $-$500 $\pm$ 20 km s$^{-1}$ and 
FWHM of 1360 $\pm$ 30 km s$^{-1}$
and a secondary wing with $V_s$ of
$-$2250 $\pm$ 30 km s$^{-1}$ and FWHM of 1110 $\pm$ 60 km s$^{-1}$.
The high [O III]/H$\beta$ flux
ratio implies that the fast outflowing gas is
ionized by an AGN. The time-averaged mass-loss rate and kinetic
power are estimated to be
160 $\mathrm{M_\odot}$ yr$^{-1}$ and $10^{44.3}$ erg s$^{-1}$, respectively. 

\section{Radio Surveys} \label{sec:obs}

\subsection{NRAO Very Large Array Survey (VLASS)}

We have developed our program to search for radio continuum emission from local ($z<$ 1) analogs of the LRDs
as follows. A first step involves looking for relatively bright ($>$ 0.5 mJy) sources in the NRAO Very Large Array Sky Survey
\citep[VLASS;][]{2020PASP..132c5001L,Kimball2026VLASS}.
This survey is being made in S-band
(2$-$4 GHz, hereafter referred to as 3 GHz), covering the entire sky at Dec $> -40^\circ$. Three
epochs of observation have been completed and the fourth and last one is currently underway. To search for radio detections we did
a visual inspection of the interactive Hierarchical Progressive Survey (HiPS) images, available at \url{https://vlass-dl.nrao.edu/vlass/HiPS/MedianStack/Quicklook/}. This data format allows users to quickly browse and zoom into very large astronomical images.

J2048 was clearly detected in all four epochs available, and the Flexible Image Transport System \citep[FITS;][]{Wells1981,Pence2010}
individual images were analyzed to determine the total flux density of the associated radio source using the software package
Astronomical Image Processing System \citep[AIPS;][]{Greisen1990,Greisen2003} of NRAO.  In Table 1 we show the epoch,
telescope, survey, central frequency, bandwidth and angular resolution of the observations, as well as the flux density of the source.
The lack of significant time variability over the 8-year period
of the VLASS observations is consistent with that observed
in other LLRDs at other bands of the electromagnetic spectrum
\citep{2025arXiv251116082B}.
A final image, made by averaging the VLASS four epochs, is shown in Figure 1. This stacked VLASS
image provides the best signal-to-noise ratio (SNR$\simeq$15) to determine the parameters
of the source, which we list in Table 2. The quoted uncertainty on the radio position, $\Delta \theta_S$, is statistical
and is approximately given by \citep{1997PASP..109..166C}:

$$\rm \Delta \theta_S \simeq \theta_B/(2 \times SNR),  \eqno{(1)}$$

\noindent where $\theta_B$ is the angular size of the synthesized beam, and
SNR is the signal-to-noise ratio of the detection.
A more realistic error can be obtained
by adding in quadrature a systematic astrometric error of $0\rlap.{''}1$, appropriate for VLASS images \citep{2020PASP..132c5001L,
2021ApJS..255...30G}.

\begin{deluxetable*}{ccccccc}
%\digitalasset
\tablewidth{0pt}
\tabletypesize{\scriptsize}
\tablecaption{Surveys used in this paper \label{tab:description}}
\tablehead{
\colhead{} & \colhead{} & \colhead{} & \colhead{Central} & \colhead{}  & \colhead{}  & \colhead{Total} \\[-0.3cm ]
\colhead{} & \colhead{} & \colhead{} & \colhead{Frequency} &  \colhead{Bandwidth} & \colhead{Angular}  & \colhead{Flux Density} \\[-0.3cm ]
%\colhead{} & \colhead{} & \colhead{} & \colhead{} & 
%\colhead{Resolution}  & \colhead{} \\ [-0.2cm]
\colhead{Epoch} & \colhead{Telescope} &\colhead{Survey} & \colhead{(GHz)} &  \colhead{(GHz)}&  \colhead{Resolution} & \colhead{(mJy)}}
\startdata
1996-Sep-05 & VLA & NVSS & 1.40 &  0.1 & $\simeq 45{''}$  & 4.0$\pm$1.1  \\
2011-Mar-31 & VLA & FIRST & 1.40 &  0.05 & $\simeq 5{''}$  & 3.6$\pm$0.3  \\
2014-2024 & LOFAR & LoTSS & 0.144 & 0.048 & $\simeq 9{''}$  & 9.7$\pm$0.9  \\
2017-Dec-16 & VLA & VLASS & 3.00 &  2.0 & $\simeq 3{''}$  & 3.0$\pm$0.3  \\
2020-Aug-05 & VLA & VLASS & 3.00 & 2.0 &  $\simeq 3{''}$  & 3.0$\pm$0.4  \\
2020-Oct-16 & ASKAP & RACS-low & 0.88 &  0.288 & $\simeq 15{''}$  & 4.6$\pm$0.7  \\
2021-Jan-17 & ASKAP & RACS-mid & 1.37 & 0.144 & $\simeq 10{''}$  & 4.0$\pm$0.3  \\
2022-Jan-10 & ASKAP & RACS-high & 1.67 & 0.288 & $\simeq 7{''}$  & 3.7$\pm$0.3  \\
2023-Jan-21 & VLA & VLASS & 3.00 & 2.0 & $\simeq 3{''}$  & 3.2$\pm$0.3  \\
2025-Sep-18 & VLA & VLASS & 3.00 & 2.0 & $\simeq 3{''}$  & 3.1$\pm$0.3  \\
\enddata
%\tablecomments{}
\tablecomments{Each mosaic of LoTSS was built from multiple runs and no single epoch can be given per source.}
\label{tab1}
\end{deluxetable*}

\begin{table}[h]
\centering
\caption{Parameters of the source from the averaged VLASS image}
\begin{tabular}{cc}
\hline
%Column 1 & Column 2  \\
\hline
Parameter & J2048 \\
\hline
RA(J2000) & $20^h~ 48^m~ 37\rlap.^s245\pm0\rlap.^s002$ \\
Dec(J2000) &  $-00^\circ~ 24'~ 37\rlap.{''}32\pm0\rlap.{''}04$ \\
Central Frequency (GHz) & 3.0 \\
Bandwidth (GHz) & 2.0 \\
Total Flux Density (mJy) & 3.0$\pm$0.2  \\
Angular Dimension ($''$) & $\leq 0.8$ \\
Proper Size (kpc) & $\leq$ 4.7  \\
Brightness Temperature (K) & $\geq$900   \\
\hline
\end{tabular}

\end{table}

The upper limit to the angular size of the source was estimated using \citep{1997PASP..109..166C}:

$$\rm \theta_S \leq \theta_B/\sqrt{SNR},  \eqno{(2)}$$

\noindent where $\theta_S$ is the angular size of the source. Using the
Javascript calculator from \citet{2006PASP..118.1711W} and
assuming a flat Planck $\Lambda$CDM cosmology with
$H_0 = 67.4\,\mathrm{km\,s^{-1}\,Mpc^{-1}}$,
$\Omega_{\rm m} = 0.315$, and
$\Omega_\Lambda = 0.685$ \citep{2020A&A...641A...6P},
we find that, for a source at $z=$ 0.4332, the angular diameter distance is $\mathrm{D_A}$ =
1202 Mpc and the luminosity distance is $\mathrm{D_L}$ = 2468 Mpc. The proper size of the source is given by $\mathrm{l = D_A \theta_S}$, 
which results in an
upper limit to the proper size of the source of $\leq$ 4.7 kpc. This upper limit is not stringent, but it will be greatly improved with the future VLBA observations.

The rest-frame brightness temperature of a source in Kelvin is given by \citep{2016era..book.....C}:

$$T_b =  1.22 \times 10^3 (1 + z)  \Biggl[{{S_\nu} \over {\rm mJy}}\Biggr] \Biggl[{{\nu} \over {\rm GHz}}\Biggr]^{-2}
\Biggl[{{\theta_S} \over {''}}\Biggr]^{-2}, \eqno{(3)} $$

\noindent which gives $T_b \geq$ 900 K for J2048. As in the case of the proper size of the
source, the lower limit to the rest-frame brightness temperature is not stringent, and higher angular resolution observations are required to improve it.
These future observations can determine \rm the rest-frame brightness temperature, a useful parameter to restrict the nature of
the radio emission. A lower limit of about 10$^6$ K is needed to strongly favor a compact non-thermal AGN component 
\citep{Condon1992,Morabito2022}.

\begin{figure}
    \centering
    \vskip-3.0cm
    \includegraphics[width=0.7\textwidth]{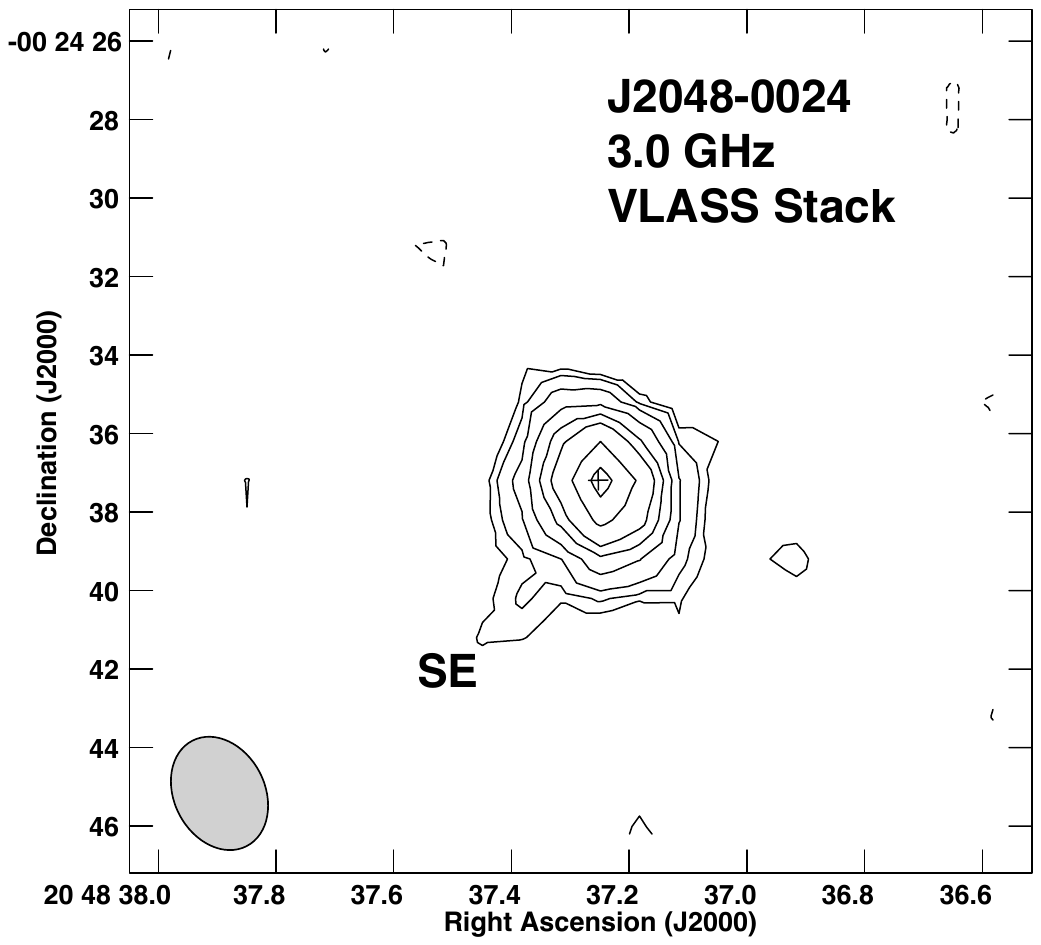}
   \vskip-3.5cm    
    \caption{\scriptsize VLA contour 3.0 GHz image of the J2048-0024 region, made averaging the four epochs available from the
Very Large Array Sky Survey (VLASS).
        Contours are -4, -3, 3, 4, 6, 10, 15, 20, 30, and 40 times
      80.0 $\mu$Jy beam$^{-1}$, i.e.,
      the rms noise in this region of the image.
%      The synthesized beam ($3\rlap.{''}60 \times 2\rlap.{''}29; +39^\circ$)
%      is shown in the bottom left corner.
      The cross marks the optical position of the source from the
      Gaia Early Data Release 3 \citep{2020yCat.1350....0G}. The morphology of the source consists of an unresolved
      component with a marginal protuberance about 4$''$ to the SE of the main component, which may be tracing a kpc jet or a second component in
      a merging ULIRG system. The tip of this protuberance is indicated with an SE label. The half-power full width of the
      beam ($2\rlap.{''}98 \times 2\rlap.{''}28; PA = +27\rlap.^\circ2$) is shown in the bottom left corner.}
    \label{fig:J2048}
\end{figure}

\subsection{Other Surveys}

J2048 has been detected in several other surveys: the NRAO VLA Sky Survey Catalog \citep[NVSS;][]{1998AJ....115.1693C}, the
Faint Images of the Radio Sky at Twenty-cm
\citep[FIRST;][]{1995ApJ...450..559B}, the LOFAR Two-metre Sky Survey \citep[LoTSS;][]{2026A&A...707A.198S}, and the Rapid
Australian Square Kilometre Array Pathfinder Continuum Survey
\citep[RACS;][]{2020PASA...37...48M}. The parameters of these observations are given in
Table 1. We have used these flux densities, as well as the one derived from the average image from all four VLASS
observations, to determine $\alpha$, the spectral index of the source, using the convention
$S_\nu \propto \nu^{\alpha}$. In Figure 2, we show the flux densities as a function of frequency,
and the least-squares fit to the data, given by 

$$\rm {log[S_\nu(mJy)] = (0.65\pm0.02) -(0.39\pm0.04)~log[\nu(GHz)]}. \eqno{(4)} $$

The spectral index $\alpha =$ -0.39$\pm$0.04 is consistent with moderately optically-thick synchrotron emission,
and rules out blackbody emission or dust emission, where spectral indices equal to or larger than 2 are expected. 
Its relative flatness suggests a potential blazar.

\begin{figure}
    \centering
    \vskip-2.0cm
    \includegraphics[width=0.5\textwidth]{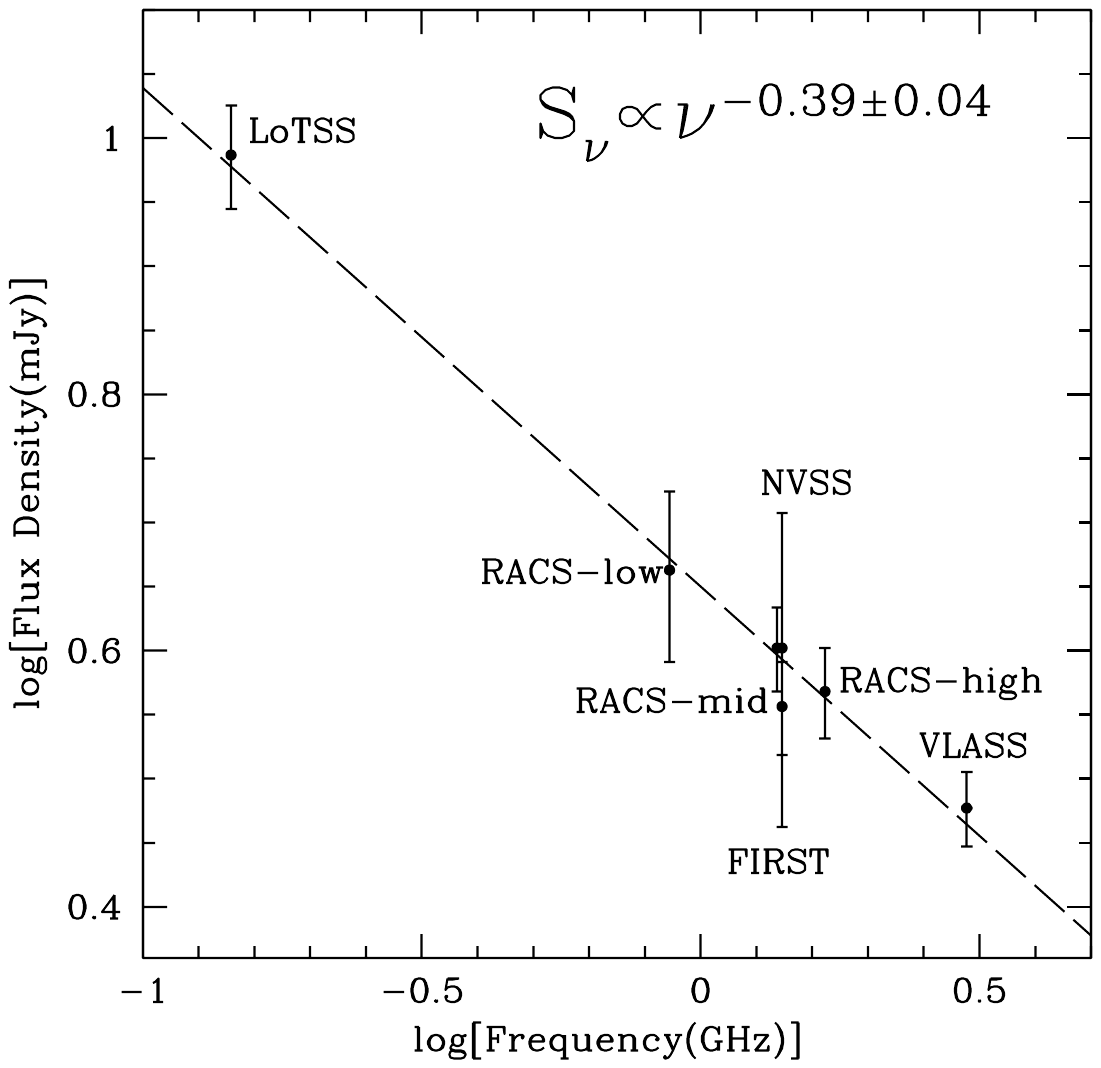}
   \vskip-2.5cm    
    \caption{\scriptsize J2048 flux density as a function of frequency. The dashed line indicates the
    least-squares fit to the data, given in the text. The data points from the different surveys used are labeled.
    The spectral index of the fit is also given in the top of the figure.}
    \label{fig:J2048SpIn}
\end{figure}

\section{The Radio Luminosity of J2048}

The radio luminosity $\mathrm{L_{\rm R}}$ between the rest frequencies $\mathrm{\nu_2}$ and $\mathrm{\nu_1}$
of
a cosmological source at a given $z$ is calculated using
the following equation:

$${L_{\rm R \it} = 4 \pi ~ D_ {\rm L \it}^2 ~ S_\nu ~ (1+z)^{-(\alpha+1)} \nu^{-\alpha} (\nu_2^{\alpha + 1}} -
\nu_1^{\alpha + 1})/(\alpha + 1). \eqno{(5)}$$

\noindent The radio luminosity has been estimated by assuming that the
power-law spectrum, $\mathrm{S_\nu \propto \nu^{\alpha}}$,
is valid between 1 and 10 GHz.
%, that is that
% $\nu_2$ = 10 GHz and $\nu_1$ = 1 GHz.
In this frequency
range, spectral curvature
is typically minimal \citep{1992ARA&A..30..575C}.
For the parameters of J2048,
we obtain $L_{\rm R} 
 = 1.2 \times 10^{41}$ erg s$^{-1}$, two orders of magnitude
larger than the radio luminosity of J1047+0739, the 
$z$ = 0.1682 LLRD analog detected in the radio
by \citet{2026A&A...707L..17R}.
The radio luminosity of J2048 places it in the lower end of the
range defined by radio-loud giant elliptical galaxies and quasars,
$10^{41-46}$ erg s$^{-1}$ \citep{1999qagn.book.....K}. Following \citet{2016ApJ...831..168K}
we refer as radio-loud to those sources that exceed a monochromatic luminosity
of $\mathrm{L_{5GHz} \simeq 10^{30.2}~erg~s^{-1}~Hz^{-1}}$ or a radio luminosity of $\mathrm{L_{R} \simeq 10^{40}~erg~s^{-1}}$.
Under these definitions, J2048 can be considered a radio-loud source.

We can use the measured flux density of J2048 to predict the flux density
expected for an identical source located at cosmological distances.
For a radio source with a power-law spectrum given by
${S_\nu \propto \nu^\alpha}$,
the flux density measured at the same frequency and for two different $z$ values is given by:
$$\Biggl[{{S_\nu(z_1)} \over {S_\nu(z_0)}} \Biggr] = \Biggl[{{1 + z_0} \over {1 + z_1}} \Biggr]^{-(\alpha+1)} ~~ \Biggl[{{D_{\rm L} (z_0)}
\over {D_{\rm L} (z_1)}} \Biggr]^2, \eqno{(6)}$$

\noindent where $z$ is the redshift, $S_{\rm \nu}$ is the flux density, and
$\mathrm{D_{\rm L}}$ is the
luminosity distance. In our case, the reference is J2048.
The flux density for a source equal to J2048,
located where the number density of LRDs peaks, i.e., $z$=5 ($\mathrm{D_{\rm L}}$ = 47660 Mpc), is estimated by

$${S_\nu(z = 5) = 7.0 \times 10^{-3} ~~ S_\nu(z = 0.4332).} \eqno{(7)}$$

The expected flux density at 3.0 GHz is, then, $\simeq$20 $\mu$Jy. This flux density
can be detected with the VLA at the 10$\sigma$ level with an on-source integration time of $\simeq$5 hours.
With the next generation VLA \citep[ngVLA;][]{Selina2018}
the on-source time required will be only about one minute.
Realistically, taking into account slewing and calibration, a time in the tens of minutes will be required.  For the reliable imaging of a resolved structure a proper \sl (u,v) \rm coverage is necessary, prolonging even further the required time.
At the representative frequency of the Square Kilometer Array-low
\citep[SKA-low;][]{Braun2020} of 200 MHz we expect a flux density of $\simeq$60 $\mu$Jy. This flux
density will be detected with the completed SKA-low at the 10$\sigma$ level with an on-source integration time of $\simeq$40 minutes.
Then, despite the higher flux density at lower frequencies, the better sensitivity of the ngVLA indicates this is the instrument of choice
for future searches.

We note that the expected 3.0 GHz flux density for an identical source to J2048 located at $z$=5 of $\simeq$20 $\mu$Jy
is very similar to that measured for PRIMER-COS 3866 ($z$ = 4.66) of $24.6\pm 4.2~ \mu$Jy \citep{Smolcic2017},
suggesting that these sources are comparable in this respect. However, these sources have different spectral indices,
$-0.39 \pm 0.04$ for J2048 and $-0.76 \pm 0.10$ for PRIMER-COS 3866 \citet{2025ApJ...986..130G}.

%The issue is how to identify an LRD similar to J2048, located at cosmological distances.

\section{ABSENCE/SUPPRESSION OF X-RAY AND RADIO EMISSION IN LRD MODELS}

Recently, models for the LRDs based on the concept of supermassive 
stars (SMSs) have been proposed \citep{2026ApJ...998..124N}. 
%On one hand, as a possible explanation for the LRDs is a 
%supermassive star \citep[SMS;][]{2026ApJ...998..124N}. 
These sources are posited to be
metal-free gigantic stars powered by nuclear fusion. Their mass
and radius can reach values of $\sim 10^{6}~M_\odot$ and 
$\sim 10^{14}$ cm, respectively, while their surface
temperature is expected to be $\sim 1.5 \times 10^4$ K
\citep{2026ApJ...998..124N}.
The monochromatic
luminosity at a rest wavelength of 4050 \AA ~is
predicted to be 
$L_\lambda \sim 1.7 \times 10^{44}$ erg~s$^{-1}~\mu$m$^{-1}$,
a value consistent with those measured in luminous LRDs, such as
MoM-BH$^*$-1 \citep{2025arXiv250316596N} and The Cliff
\citep{2025A&A...701A.168D}. No X-ray or radio emission is expected from an SMS. Indeed, the detection of X-ray emission from
the LRD 3DHST-AEGIS-12014 (renamed as
the “X-ray dot” or XRD), a compact source at $z$ = 3.28
\citep{2026ApJ..1000L..18H}, can be taken to weigh against the 
SMS hypothesis.

On the other hand, we have the black hole star 
\citep[BH*, also called a quasi-star;][]{2025arXiv250316596N, 2026ApJ...996...48B} model, which 
proposes an extraordinary object: an accreting supermassive black hole
surrounded by a massive envelope of gas. In contrast to the SMSs, 
which derive their luminosity from nuclear fusion,
the BH*s are heated by the energy released by the accreting disk
of the central supermassive black hole. The existence of these
remarkable objects was already speculated in papers written
one to two decades ago \citep{2006MNRAS.370..289B, 2008MNRAS.387.1649B, 2011MNRAS.414.2751B, 2013ApJ...778..178H}.
The model by \citet{2026ApJ...998L...4S} proposes that
a BH* has characteristic total mass, radius, temperature,
luminosity and lifetime of $\sim$ $10^6~\mathrm{M_\odot}$, $\sim$$2.4\times10^5~\mathrm{R_\odot}$, $\sim$5,000 K,
$\sim 3.6 \times 10^{10}~L_\odot$ and $\sim$20 Myr, respectively.
The BH* model strongly inhibits the X-ray and radio emissions,
the first by the Compton opacity and the second by the free-free
opacity. However, in the case of the radio emission (assumed to come from the jets from the central black hole), the jets can escape from the ionized envelope and become detectable in the radio. In the case of J2048 high angular resolution radio observations will help to establish if we are seeing the base of a relativistic jet or more extended lobes that could have escaped from the envelope of a BH*. These observations are in progress.

In the BH* model, the initial mass of the central black hole can be as small as $\sim$10~$M_\odot$, but it grows rapidly, in a scale of
around $10^6$ yr, reaching a million 
times the mass of the Sun at the end of its lifetime. The
SMBH formed at the end of the
life of the black-hole star could act as a massive seed for further
growth and may help understand the 
existence of $\sim$$10^9~\mathrm{M_\odot}$ SMBHs early ($\sim$1 Gyr)
in the universe.

\section{The radio observations of J2048 in a multiwavelength context}

The radio source in J2048 has a spectral index of -0.39, which likely corresponds to a jet with partially optically thick radio emission. The main component has an angular diameter of less than 0.8 arcsec. At $z =$ 0.4332, this upper limit to the angular diameter corresponds to an upper limit to the proper size of 4.7 kpc. In addition, the second, fainter component to the SE is displaced by $\simeq$3.6 arcsec from the center of the main component. This angular displacement corresponds to a physical distance in the plane of the sky of 21.0 kpc. 
From the comparison of the $\leq$4.7 kpc main component of the jet with Fig. 2 of 
\citet{2026ApJ...999...30C},
it is inferred that such main component is embedded in and surrounded by a more extended region, of up to $r$ = 7 kpc, of young stellar UV continuum, narrow 
H$\alpha$, and outflow [O III] line emissions. This suggests that the main component of the radio jet in J2048 could be constrained in length and luminosity by inverse
Compton scattering of the UV photons by the highly relativistic electrons that would otherwise
produce the synchrotron emission, a mechanism proposed in
\citet{2022NewAR..9401642M} for BH-jets associated with 
high-mass star formation environments. 

%If the SMBH in J2048 is as massive as $10^{10.2} M_\odot$ 
%\citep{2026ApJ...999...30C},
The radio flux density of 3-4 mJy at $z =$ 0.4332 implies an intrinsic radio luminosity of $1.2 \times 10^{41}$ erg s$^{-1}$ between 1 and 10 GHz. 
%However, the radio luminosity of similarly massive SMBHs accreting at the Eddington rate is $\sim 10^{43}$ erg s$^{-1}$ \citep{2003MNRAS.345.1057M}, about two orders of magnitude larger. This suggests a possible
%suppression by two orders of magnitude of the luminosity of the relativistic radio jet. 
What is the expected radio luminosity for a source like J2048?

From the fundamental-plane relation of \citet{2003MNRAS.345.1057M}
we have that

%\begin{equation}
$$
\log L_{5GHz} =
0.60\,\log L_X
+0.78\,\log\left(\frac{M_{\rm BH}}{M_\odot}\right)
+7.33, \eqno(8) $$
%\end{equation}

\noindent where $L_{5GHz}$ is the total radio luminosity derived from observations at 5 GHz and $L_X$ is the
2–10 keV luminosity, both given in erg s$^{-1}$. $M_{\rm BH}$ is the mass of the black hole given in solar masses.
\citet{2003MNRAS.345.1057M} note that log($L_{5GHz}$) has a substantial scatter of $\sigma_{5GHz}$ = 0.88.

The X-ray bolometric correction, $\kappa_X$ is defined as

%\begin{equation}
$$
\kappa_X \equiv \frac{L_{\rm bol}}{L_X},  \eqno(9)$$
%\end{equation}

\noindent where $L_{\rm bol}$ is the bolometric luminosity of the source.
If the black hole is radiating at the Eddington luminosity, we can assume that

%\begin{equation}
$$
L_{\rm bol}=L_{\rm Edd}
=1.26\times10^{38}
\left(\frac{M_{\rm BH}}{M_\odot}\right)
{\rm erg\,s^{-1}}. \eqno(10)
$$
%\end{equation}

Therefore,

%\begin{equation}
$$
L_X =
\frac{1.26\times10^{38}}{\kappa_X}
\left(\frac{M_{\rm BH}}{M_\odot}\right)
{\rm erg\,s^{-1}}. \eqno(11) $$
%\end{equation}

Substituting this expression into the fundamental-plane relation gives

%\begin{equation}
%\boxed{
$$
\log L_{5GHz} =
30.19
+1.38\log\left(\frac{M_{\rm BH}}{M_\odot}\right)
-0.60\log(\kappa_X). \eqno(12) $$
%}.
%\end{equation}

J2048 has $L_{\rm bol} = 10^{13.6~L_\odot}$, which following \citet{Duras2020}
implies $\kappa_X \simeq 80$. Finally, since $M_{\rm BH} = 10^{10.2}~M_\odot$,
we obtain $L_{5GHz} \simeq 10^{43.1 \pm 0.9}~erg~s^{-1}$
We note, however, that \citet{2026ApJ...999...30C}
report  an Eddington ratio of 0.08$\pm$0.01 for J2048, so the expected value will be $L_{5GHz} \simeq 10^{42.0 \pm 0.9}~erg~s^{-1}$.
This radio luminosity is about one order of magnitude larger than that measured, suggesting that a suppression mechanism could be
present. However, the scatter in the \citet{2003MNRAS.345.1057M} relation is also about an order of magnitude, so the 
smaller radio luminosity could be simply due to the source being intrinsically weaker.

The suppressed radio emission of the jets, if present, should be expected to
produce an inefficient radio emitting massive outflow. In fact, J2048 hosts a massive ionized outflow 
with extension of up to 3 kpc, a velocity of 2070$\pm$ 40 km s$^{-1}$, a time-averaged mass-loss rate of 160 $M_\odot$ yr$^{-1}$,  an outflow timescale of 0.7 Myr, and a kinetic power of $10^{44.3}$ erg s$^{-1}$ 
\citep{2026ApJ...999...30C}. These authors propose that this outflow must be driven by AGN activity because of its high velocity, kinetic power, and AGN-type ionization. 
%However, thanks to the relatively close distance of J2048, the idea that the extended blue continuum is emitted by young stars rather than scattered AGN %light is strongly supported. 
They also found that the  massive ionized outflow in J2048 is only marginally resolved, and unfortunately no 
preferred position angle was reported to compare with the radio morphology.

There are many evidences that, in LRD analogs, the concurrency of nuclear starbursts with an SMBH fed by dense cold gas could be common \citep[e.g. ][]{2024ApJ...968....4P, 2025ApJ...992...71R, 2025A&A...704A.313D, 2026ApJ..1000...90I, 2026arXiv260416178E}. 
J2048 is an ULIRG with an infrared luminosity (1–1000 $\mu$m) higher than $10^{12}~ \mathrm{L_\odot}$. In the local universe, there are advanced mergers of massive cold gas-rich galaxies with their respective nuclei closer than 10 kpc 
\citep{1996ARA&A..34..749S}. \citet{1988ApJ...328L..35S} proposed that ULIRGs are the initial dust-enshrouded stage of quasars. They have high accretion rates of cold gas that feed both a nuclear SMBH and a starburst, as believed to be the case of LRDs in the early universe \citep{2026ApJ..1000...90I}. This is why the ULIRG J2048 qualitatively reflects the properties of high-$z$ LRDs: it has a V-shaped SED, a blue continuum at rest UV wavelengths, a red continuum at rest optical wavelengths, and a large $\mathrm{M_{\rm BH}/M_*}$ ratio of 60\%. It is an analog of high-$z$ LRDs in a late stage of evolution, when the merging process induces a nuclear starburst of 
$400\pm 60~\mathrm{M_\odot}$ yr$^{-1}$, and an X-ray-obscured AGN with a bright radio jet.  

Close to the SMBH the LRDs will go radio-undetected  because of the extremely large free-free opacity of the BH ionized envelope. 
From the parameters of the model of  \citet{2026ApJ...998L...4S} presented in Section 5 we estimate a characteristic
electron density of $\mathrm{n_e \simeq 6.0 \times 10^{13}~cm^{-3}}$ and an emission
measure of $\mathrm{EM \simeq 2.0 \times 10^{25}~\rm pc\,cm^{-6}}$. From the expression for free-free opacity \citep{Wilson2022}:

$$\tau_{\rm ff} \simeq
3.28 \times 10^{-7}
\left(\frac{T_e}{10^4\,{\rm K}}\right)^{-1.35}
\left(\frac{\nu}{\rm GHz}\right)^{-2.1}
\left(\frac{\rm EM}{\rm pc~\,cm^{-6}}\right), \eqno{(13)}$$

\noindent we obtain an enormous opacity, $\tau_{\rm ff} \simeq
1.7 \times 10^{18}$ at 3.0 GHz. Radio emission from the central regions will not escape the envelope.

Outside of this envelope the jets may be silenced at radio due to the inverse Compton scattering by photons from the ionized
envelope  on the jet highly relativistic electrons that otherwise
would produce the synchrotron emission. To explore this possibility we note that the energy losses of the relativistic electrons to synchrotron 
emission and IC scattering are proportional to the energy densities of the magnetic and radiation fields, respectively.
From \citet{Lobanov1998} and \citet{OSullivan2009} we find that at the surface of the envelope, at a distance of 0.005 pc from the AGN,
a magnetic field of order $10^2$ G is expected. The magnetic energy density at this location is then $\mathrm{u_B = B^2/(8~\pi) \simeq 4.0 \times 10^2~erg~cm^{-3}}$. For a spherical surface with radius $R$ emitting a luminosity $L$, the radiation energy density will be given by
$\mathrm{u_{rad} = L/(4 \pi R^2 c)}$, where $c$ is the speed of light. For $\mathrm{L \simeq 3.6 \times 10^{10}~L_\odot = 1.4 \times 10^{44}~erg~s^{-1}}$,
and $\mathrm{R \simeq 1.7 \times 10^{16} ~cm}$ we obtain $\mathrm{u_{rad}\simeq 1.3~erg~cm^{-3}}$. Since $\mathrm{u_B/u_{rad} \simeq 300}$ we conclude that IC cooling is unimportant in comparison to synchrotron cooling and will not significantly suppress the synchrotron emission.

%If the extended hosts of LRDs are nuclear bursts of massive stars, the LRD BH-jets may be naturally radio silent, close to the SMBH because
%the large free-free opacity of the ionized envelope
%and outside of this envelope due to the microscopic inverse
%Compton scattering by UV photons from massive stars on the highly relativistic electrons that produce the synchrotron emission of BH-jets.

\section{Conclusions}

Our conclusions can be summarized as follows.

    1) The radio emission of J2048 at 3.0 GHz  lacks significant time variability over an 8-year period of observations. This is consistent with the lack of significant time variability in observations of other LLRDs at different bands of the electromagnetic spectrum \citep{2025arXiv251116082B}.
The spectral index $\alpha$ = -0.39$\pm$0.04 of the radio emission is consistent with moderately optically-thick synchrotron emission, and rules out blackbody emission or dust emission. The spectrum flatness could indicate a potential blazar. 
 
    2) The upper limit to the
    diameter of the main radio component of the source is $\leq$4.7 kpc.  The properties of this main radio component will be studied in detail by already proposed VLBA observations. The radio source in J2048 has a second component to the SE of the main component, with a total physical separation in the plane of the sky of 21.0 kpc. This second component could be the tip of an SMBH-jet escaping to the intergalactic medium, from the nuclear high densities of gas and radiation. 

    3) The radio luminosity of J2048 of 
    $\mathrm{1.2 \times 10^{41} ~erg ~s^{-1}}$ place it in the lower end of the range defined by radio-loud giant elliptical galaxies and quasars of 
    $\mathrm{10^{41-46} ~erg ~s^{-1}}$ \citep{1999qagn.book.....K}. The flux density at 3.0 GHz of a source like J2048 at $z$=5 is $\simeq$20 $\mu$Jy. This flux density can be detected with the VLA at the 10$\sigma$ level with an on-source integration time of $\simeq$5 hours. The ngVLA will achieve such a detection in a matter
    of minutes. 
%    The problem is how to identify such a target. 
    
    4) The radio luminosity of the SMBH of $\mathrm{10^{10.2} ~ M_\odot}$ in J2048 \citep{2026ApJ...999...30C} 
 is expected to be about an order of magnitude larger than measured, 
 suggesting that its radio luminosity could potentially be strongly dimmed.  The suppressed energy of the relativistic electrons in the jets could convert
    them into an inefficient radio-synchrotron emitting massive outflow. In fact, J2048 hosts a massive ionized outflow with extension of up to 3 kpc with a kinetic power of 
    $\mathrm{10^{44.3} ~erg ~s^{-1}}$ \citep{2026ApJ...999...30C}.

 5) J2048 appears to be an ULIRG with an infrared luminosity (1–1000 µm) of at least $\mathrm{10^{12} ~L_\odot}$. J2048 and ULIRGs in general may reflect the properties of cosmological LRDs because in both types of galaxies dense cold gas feeds at high rates nuclear starbursts and the growing of SMBHs. 
 
%1. We report the radio detection of the LLRD J2048. Its radio
%luminosity is $L_{\rm R} = 1.2 \times 10^{41}$ erg s$^{-1}$,
%in the lower end of the
%range defined by radio-loud giant elliptical galaxies and quasars, %i.e.,
%$10^{41-46}$ erg s$^{-1}$.

%2. This detection confirms that LLRDs can be detected in the radio,
%and suggests that selected LRDs could be detected with 
%moderately long (hours) integrations made with the VLA.

%3. We discuss several mechanisms that could restrain the
%radio emission from J2048.

%% Please use the acknowledgment and contribution environments. This will
%% be anonomyized when the "anonymous" style option is used.
\begin{acknowledgments}
We thank an anonymous referee for valuable suggestions that improved the paper.
This research has made use of the SIMBAD database, operated at CDS, Strasbourg, France. LFR thanks the financial support
from grant CBF-2025-I-2471 of SECIHTI, México. 
RAGL acknowledges the financial support of DGAPA, UNAM,
project IN106124.
LL acknowledges the support of DGAPA-PAPIIT grant IN108324 and SECIHTI grant CBF-2025-I-109.

\end{acknowledgments}

%\begin{contribution}
%LFR was responsible for interpreting the data and for writing and submitting the manuscript.

%%This section gives authors the space to recognize author contributions. The text inside this environment is NOT counted towards the total word quanta. At a minimum, manuscripts are expected to include this text:

%All authors contributed equally to the Terra Mater collaboration.

%% But authors are expected to provide more specific details, e.g.
%%
%%SC was responsible for writing and submitting the manuscript.
%%WWM came up with the initial research concept and edited the manuscript.
%%OTS obtained the funding and edited the manuscript.
%%EBF provided the formal analysis and validation. He also edited the manuscript.
%%GEH Supervised the undergraduates, wrote the software and administers the project github and Zenodo repositories.
%%
%% Authors can use the Contributor Role Taxonomy (CRediT) at
%% https://credit.niso.org
%% for ideas on how write a good statement tailored to their needs.

%\end{contribution}
%% To help institutions obtain information on the effectiveness of their
%% telescopes the AAS Journals has created a group of keywords for telescope
%% facilities.
%
%% Following the acknowledgments section, use the following syntax and the
%% \facility{} or \facilities{} macros to list the keywords of facilities used
%% in the research for the paper.  Each keyword is check against the master
%% list during copy editing.  Individual instruments can be provided in
%% parentheses, after the keyword, but they are not verified.
\facilities{NRAO VLA, CSIRO ASKAP}

%% Similar to \facility{}, there is the optional \software command to allow
%% authors a place to specify which programs were used during the creation of
%% the manuscript. Authors should list each code and include either a
%% citation or url to the code inside ()s when available.
\software{astropy \citep{2013A&A...558A..33A,2018AJ....156..123A,2022ApJ...935..167A}.
%          Cloudy \citep{2013RMxAA..49..137F},
 %         Source Extractor \citep{1996A&AS..117..393B}
          }

%% Appendix material should be preceded with a single \appendix command.
%% There should be a \section command for each appendix. Mark appendix
%% subsections with the same markup you use in the main body of the paper.
%%
%% Each Appendix (indicated with \section) will be lettered A, B, C, etc.
%% The equation counter will reset when it encounters the \appendix
%% command and will number appendix equations (A1), (A2), etc. The
%% Figure and Table counter will not reset.

\bibliography{J2048Lib}{}
\bibliographystyle{aasjournalv7}

%% This command is needed to show the entire author+affiliation list when
%% the collaboration and author truncation commands are used.  It has to
%% go at the end of the manuscript.
%\allauthors

%% Include this line if you are using the \added, \replaced, \deleted
%% commands to see a summary list of all changes at the end of the article.
%\listofchanges

\end{document}